\documentclass[11pt,prd,preprint,superscriptaddress]{revtex4}
\usepackage{revsymb}
\usepackage{graphicx}
\usepackage{bm}
\usepackage{amsmath}
\usepackage{amssymb}
\usepackage{graphicx}

\def\beq{\begin{eqnarray}}
\def\eeq{\end{eqnarray}}

\def\La{\Lambda}

\begin{document}

\title{Bulk Viscous Cosmology}
\author{R. Colistete Jr.\footnote{E-mail address: rcolistete@terra.com.br}}
\affiliation{Universidade Federal do Esp\'{\i}rito Santo,
Departamento de F\'{\i}sica\\ Av. Fernando Ferrari, 514, Campus de
Goiabeiras, CEP 29075-910, Vit\'oria, Esp\'{\i}rito Santo, Brazil}

\author{J.C. Fabris\footnote{E-mail address: fabris@pq.cnpq.br}}
\affiliation{Universidade Federal do Esp\'{\i}rito Santo,
Departamento
de F\'{\i}sica\\
Av. Fernando Ferrari, 514, Campus de Goiabeiras, CEP 29075-910,
Vit\'oria, Esp\'{\i}rito Santo, Brazil}

\author{J. Tossa\footnote{E-mail address: joel.tossa@imsp-uac.org}}
\affiliation{Institut de Math\'ematiques et de Sciences Physiques
- IMSP\\
Universit\'e d'Abomey-Calavi, BP613, Porto Novo, B\'enin}

\author{W. Zimdahl\footnote{E-mail: zimdahl@thp.uni-koeln.de}}
\affiliation{Universidade Federal do Esp\'{\i}rito Santo,
Departamento
de F\'{\i}sica\\
Av. Fernando Ferrari, 514, Campus de Goiabeiras, CEP 29075-910,
Vit\'oria, Esp\'{\i}rito Santo, Brazil}

\begin{abstract}
We propose a scenario in which the dark components of the Universe
are manifestations of a single bulk viscous fluid. Using dynamical
system methods, a qualitative study of the homogeneous, isotropic
background scenario is performed in order to determine the phase
space of all possible solutions. The specific model which we
investigate shares similarities with a generalized Chaplygin gas
in the background but is characterized by non-adiabatic pressure
perturbations.  This model is tested against supernova type Ia and
matter power spectrum data. Different from other unified
descriptions of dark matter and dark energy, the matter power
spectrum is well behaved, i.e., there are no instabilities or
oscillations on small perturbation scales. The model is
competitive in comparison with the currently most popular
proposals for the description of the cosmological dark sector.
\end{abstract}


\maketitle
\date{\today}

\section{Introduction}

The crossing of observational data from high redshift supernovae
of type Ia (SNe Ia) \cite{SN,tonry,riess,astier}, cosmic microwave
background (CMB) radiation \cite{spergel}, matter power spectra
\cite{tegmark}, X-rays from clusters of galaxies
\cite{allen,linder}, and weak gravitational lensing
\cite{challinor}  strongly suggests that the present Universe is
dynamically dominated by a dark sector which is responsible for
about $96\%$ of its total energy content. It is usually assumed
that this dark sector has two different components: (i) dark
matter, which is supposed to consist of weakly interacting massive
particles (WIMPS) with zero effective pressure and (ii) dark
energy, a mysterious entity which is equipped with a negative
pressure. Dark matter candidates include axions (a particle
present in the multiplet of grand unified theories) and
neutralinos (light particles present in broken supersymmetric
models), but none of these particles could be detected until now.
The most natural dark energy candidate is a cosmological constant
which arises as the result of a combination of quantum field
theory and general relativity. However, its theoretical value is
between 60-120 orders of magnitude greater than the observed value
for the dark energy. An alternative to the cosmological constant
is a self-interacting scalar field, known as quintesssence. For a
brief but enlightening review of these and other proposals, see
\cite{hannestad} and references therein.
\par
There exists another route of investigations in which dark matter
and dark energy are described within a one-component model.
According to this idea, dark matter and dark energy are just
``different faces" of a single, exotic fluid. To the best of our
knowledge, the first proposal along this line was the Chaplygin
gas in its original and modified forms
\cite{pasquier,fabris1,bertolami}. However, this unified
description of dark energy and dark matter, in spite of many
attractive features, seemed to suffer from a major drawback: it
predicted strong small scale oscillations or instabilities  in the
matter power spectrum, in complete disagreement with the
observational data \cite{tegmarkbis}. (On the other hand, the
observed matter power spectrum corresponds to the baryonic matter
distribution which does not exhibit strong oscillations
\cite{hermano}, so that this point is still controversial.) The
apparently unrealistic predictions of the unified Chaplygin gas
type models are the result of an adiabatic perturbation analysis.
It has been suggested that non-adiabatic perturbations may
alleviate or even avoid this problem\cite{NJP,ioav,zimdahl}.

This paper explores to what extent a viscous fluid can provide a
unified description of the dark sector of the cosmic medium. The
general influence of shear and bulk viscosity on the character of
cosmological evolution has been studied, e.g. in \cite{BeKha}, in
the context of Bianchi type I models. Under the conditions of
spatial homogeneity and isotropy, a scalar bulk viscous pressure
is the only admissible dissipative phenomenon. The cosmological
relevance of bulk viscous media has been investigated in some
detail for an inflationary phase in the early universe (see
\cite{Roy,WZ,RM,Z2nd} and references therein). However, as was
argued in \cite{antif,NJP}, an effective bulk viscous pressure can
also  play the role of an agent that drives the present
acceleration of the Universe. (Notice that the possibility of a
viscosity dominated late epoch of the Universe with accelerated
expansion was already mentioned in \cite{PadChi}). For a
homogeneous and isotropic universe, the $\Lambda$CDM model and the
(generalized) Chaplygin gas models can be reproduced as special
cases of this imperfect fluid description \cite{NJP}. Moreover, in
a gas dynamical model the existence of an effective bulk pressure
can be traced back to a non-standard self-interacting force on the
particles of the gas. While these investigations were performed
for the homogeneous and isotropic background dynamics (a study of
the background dynamics which is similar to the setup of the
present paper was recently performed in \cite{Szydlowski}), a
first perturbation theoretical analysis for a unifying viscous
fluid description of the dark sector was performed in \cite{rose}.

The bulk viscous pressure $p_{visc}$ will be described by Eckart's
expression \cite{eckart} $p_{visc} = - \xi u^{\mu}_{;\mu}$, where
the (non-negative) quantity $\xi$ is the (generally not constant)
bulk viscosity coefficient and $u^{\mu}_{;\mu}$ is the fluid
expansion scalar which in the homogeneous and isotropic background
reduces to $3 H$, where $H = \frac{\dot a}{a}$ is the Hubble
parameter and $a$ is the scale factor of the Robertson-Walker
metric. By this assumption we ignore all the problems inherent in
Eckart's approach which have been discussed and resolved within
the Israel-Stewart theory \cite{israela,israelb} (see also
\cite{Roy,WZ,RM,Z2nd} and references therein). We expect that for
the applications we have in mind here, the differences are of
minor importance.

It is obvious that the bulk viscosity contributes with a negative
term to the total pressure and hence a dissipative fluid seems to
be a potential dark energy candidate. However, a cautionary remark
is necessary here. In traditional non-equilibrium thermodynamics
the viscous pressure represents a (small) correction to the
(positive) equilibrium pressure. This is true both for the Eckart
and for the Israel-Stewart theories. Here we shall admit the
viscous pressure to be the dominating part of the pressure. This
is clearly beyond the established range of validity of
conventional non-equilibrium thermodynamics. As already mentioned,
non-standard interactions are required to support such type of
approach \cite{antif,NJP}. Of course, this reflects the
circumstance that dark energy is anything but a ``standard" fluid.
(We mention that viscosity has also been suggested to have its
origin in string landscape \cite{She}). To successfully describe
the transition to a phase of accelerated expansion, preceded by a
phase of decelerated expansion in which structures can form, it is
necessary that the viscous pressure is negligible at high
redshifts but becomes dominant later on.

In ref. \cite{rose} a one-component bulk viscous model (BV model)
of the cosmic medium was investigated in which baryons were not
taken into account. As far as the SNe Ia data are concerned, the
results of this model were similar to those obtained for a
generalized Chaplygin gas (GCG) model. But while GCG models
predict small scale instabilities and oscillations at the
perturbative level, the corresponding matter power spectrum of the
BV model turned out to be well behaved. Of course, the results of
a model that does not include baryons cannot be seen as
conclusive; after all, the observed power spectrum describes the
distribution of baryonic matter. In the present paper, we perform
an advanced analysis which properly  includes a separately
conserved baryon component. Thus we establish a more realistic
viscous fluid scenario of the cosmic substratum.

For the quantitative calculations will use a constant coefficient
of bulk viscosity. A qualitative analysis with the help of
dynamical system methods is applied to visualize the space of more
general cosmological background scenarios. The corresponding phase
space reveals interesting new features compared with those
generally found in similar models (see, for example, \cite{coley}
and references therein). In the present approach there is an
entire singular axis.  Solutions of the desired type are
generated, i.e., solutions for which an initial subluminal
expansion is followed by superluminal expansion. These solutions
are characterized by a bulk viscosity coefficient with a power law
dependence on the energy density, $\xi = \xi_{0} \rho^{\nu}$,
where $\xi_{0} = $const and $\nu < 1/2$.

We test the results of our model both against SNe Ia data and the
observed matter power spectrum. Both the results for type Ia
supernovae and for the matter power spectrum show that the BV
model is competitive with the $\Lambda$CDM model as well as with
quintessence and different Chaplygin gas type models. In
particular, our minimum $\chi ^{2}$ value for the SNe Ia data is
similar to the $\chi ^{2}$ values of those models. Furthermore,
the matter power spectrum represents a good fit to the
corresponding observational data. We argue that the absence of
oscillations and instabilities is a consequence of the fact that
the pressure perturbations in our model are intrinsically
non-adiabatic.

The paper is organized as follows. In Section \ref{background} we
define our model and specify the homogeneous and isotropic
background dynamics. Section \ref{dynamical system} performs a
qualitative analysis using dynamical system methods. In section
\ref{SN}, the SNe Ia data are used to restrict the values of the
physically relevant free parameters. In section \ref{matter power
spectrum} the matter power spectrum is determined and compared
with large scale structure observations. In section
\ref{conclusions} we present our conclusions.

\section{Background relations}
\label{background}

A bulk viscous fluid is characterized by an energy density $\rho$
and a pressure $p$ which has a conventional component $p_{\beta} =
\beta\rho$ and a bulk viscosity component $p_{visc}=-
\xi(\rho)u^\mu_{;\mu}$, such that
\begin{equation}
\label{bulk} p = \beta\rho - \xi(\rho)u^\mu_{;\mu} \quad.
\end{equation}
On thermodynamical grounds the bulk viscosity coefficient
$\xi(\rho)$ is positive, assuring that the viscosity pushes the
effective pressure towards negative values. In fact, the
expression (\ref{bulk}) is the original proposition for a
relativistic dissipative process \cite{eckart}. As already
mentioned, it follows from the more general Israel-Stewart theory
\cite{israela,israelb} in the limit of a vanishing relaxation
time. We shall assume this approximation to be valid throughout
the paper.

Let us consider the cosmic medium to consist of a viscous fluid of
the type (\ref{bulk}), which is supposed to characterize the dark
sector, and of a pressureless fluid that describes the baryon
component. Hence, the relevant set of equations is
\begin{eqnarray}
R_{\mu \nu }-\frac{1}{2}g_{\mu \nu }R &=&8\pi G\biggr\{T_{\mu \nu
}^{v}+T_{\mu \nu }^{b}\biggl\}\quad , \\
{T_{v}^{\mu \nu }}_{;\mu }=0\quad &,&\quad T_{v}^{\mu \nu }=(\rho
_{v}+p_{v})u^{\mu }u^{\nu }-p_{v}g^{\mu \nu }\quad , \\
p = p_{v} &=&\beta \rho _{v}-\xi (\rho _{v}){u^{\mu }}_{;\mu }\quad ,\label{pv} \\
{T_{b}^{\mu \nu }}_{;\mu }=0\quad &,&\quad T_{b}^{\mu \nu }=\rho
_{m}u^{\mu }u^{\nu }\quad .
\end{eqnarray}
The (super)subscripts $v$ and $b$ indicate the viscous and the
(baryonic)  matter components, respectively. Since the matter is
pressureless, the total pressure $p$ of the cosmic medium
coincides with the pressure $p_{v}$ of the viscous fluid. For the
homogeneous and isotropic background dynamics we shall restrict
ourselves to the flat Friedmann-Lema\^{\i}tre-Robertson-Walker
(FRLW) metric,
\begin{equation}
ds^{2}=dt^{2}-a^{2}(t)(dx^{2}+dy^{2}+dz^{2})\quad ,
\end{equation}
favored by the CMB anisotropy spectrum \cite{spergel}. The dynamic
equations then are:
\begin{eqnarray}
\label{em1}
\biggr(\frac{\dot a}{a}\biggl)^2 &=& \frac{8\pi G}{3}(\rho_{b} + \rho_v) \quad , \\
\label{em5}
2\frac{\ddot{a}}{a} + \frac{\dot{a}^2}{a^2}& = & - 8 \pi G p_v \\
\label{em2}
\dot\rho_v + 3\frac{\dot a}{a}\biggr(\rho_v + p_v\biggl) &=& 0 \quad , \\
\label{em3}
\dot\rho_{b} + 3\frac{\dot a}{a}\rho_{b} &=& 0 \quad , \\
\label{em4} p_v &=& \beta\rho_v - 3\frac{\dot a}{a}\xi_0\rho^\nu_v
\quad ,
\end{eqnarray}
where the bulk viscosity coefficient was assumed to have a power
law dependence on the energy density $\rho _{v}$ according to $\xi
(\rho _{v})=\xi _{0}\rho _{v}^{\nu }$ with $\xi _{0} = $ const.
The dot denotes differentiation with respect to the cosmic time.
Of course, not all these equations are independent. Equation
(\ref{em3}) is decoupled and leads to
\begin{equation}
\rho _{b}=\frac{\rho_{b0}}{a^{3}}\quad .
\end{equation}
Here and in the following, quantities with a subscript 0 refer to
the present epoch and we have used $a_{0} = 1$. From now on we
will set $\beta =0$, thus assuming the dissipative pressure to be
the dominating contribution.

With a change of variables,
\begin{equation}
\dot\rho_v = \frac{d\rho_v}{dt} = \frac{d\rho_v}{da}\frac{da}{dt}
= \rho^{\prime}_v\,\dot a \quad , \label{prime}
\end{equation}
where the prime means derivative with respect to the scale factor
$a$, and using relation (\ref{em1}), the conservation equation for
the viscous component becomes
\begin{equation}
\rho_v^{\prime}+ \frac{3}{a}\biggr(\rho_v - 3 \frac{\dot a}{a}%
\xi_0\,\rho_v^\nu\biggl) = 0 \quad .
\end{equation}
This equation can be recast in an integral form:
\begin{eqnarray}  \label{def}
\int \frac{d\varepsilon}{\varepsilon^\nu(\varepsilon + 1)} &=&
\frac{2k}{3(1 - 2\nu)}a^{\frac{3}{2}(1 -
2\nu)} \quad , \quad \nu \neq \frac{1}{2} \quad ; \\
\int \frac{d\varepsilon}{\varepsilon + 1} &=& k\,\mbox{ln}\, a
\quad , \quad \nu = \frac{1}{2}
\quad ; \\
\varepsilon = \frac{\rho_v}{\rho_{b}} \quad &,& \quad k = 9\sqrt{\frac{8\pi G}{3}}%
\xi_0\,\rho_{b0}^{\nu - \frac{1}{2}} \label{defk}\quad .
\end{eqnarray}
We will be mainly interested in the case $\nu \neq \frac{1}{2}$.

There is a simple, direct relation between $\rho _{v}$ and the scale factor $%
a$ for $\nu =0$. In this particular case, we find,
\begin{equation}
\rho _{v}=\rho _{b}\biggr\{\biggr[B+\frac{k}{3}a^{\frac{3}{2}}\biggl]^{2}-1%
\biggl\}\quad , \qquad \quad\left(\nu = 0\right)\ ,
\label{viscous-dens}
\end{equation}
where $B$ is an integration constant. It can easily be verified
that $\rho
_{v}\rightarrow a^{-3}$ when $a\rightarrow 0$ (pressureless matter) and $%
\rho _{v}\rightarrow $ cte (cosmological constant) when
$a\rightarrow \infty $. It is expedient to notice that for $\nu =
0$ the total energy density $\rho =\rho _{v}+\rho _{b}$ coincides
with the energy density of a specific GCG \cite
{pasquier,fabris1,bertolami}. Generally, a GCG is characterized by
an equation of state ($E= $ const $> 0$)
\begin{equation}
p_{\mathrm{GCG}} = - \frac{E}{\rho_{\mathrm{GCG}}^{\alpha}}\ ,
\label{pGCG}
\end{equation}
which corresponds to an energy density
\begin{equation}
\rho_{\mathrm{GCG}}=\left[E+\frac{F}{a^{3\,(1+\alpha)}}\right]^{{1}/{\left(1+\alpha\right)}}\
, \label{rhoGCG}
\end{equation}
where $F$ is another (non-negative) constant. With the
identifications $E = \frac{k}{3}\rho_{b0}^{1/2}$ and $F = B
\rho_{b0}^{1/2}$ it becomes obvious that for $\alpha =-1/2$ the
energy densities $\rho =\rho _{v}+\rho _{b}$ from
(\ref{viscous-dens}) and $\rho_{\mathrm{GCG}}$ in (\ref{rhoGCG})
coincide. The analogy between dissipative and GCG models was also
pointed out in \cite{Szydlowski}.

The structure of the total energy density $\rho =\rho _{v}+\rho
_{b}$ allows us to perform a decomposition of the viscous-baryon
system into three non-interacting components plus the baryonic
fluid:
\begin{equation}
\rho =\rho _{1}+\rho _{2}+\rho _{3}+\rho _{b}\quad ,
\label{decomp}
\end{equation}
with
\begin{equation}
\rho _{1}= E^{2}\quad ,\quad \rho _{2}=2\frac{E F}{a^{\frac{3%
}{2}}}\quad ,\quad \rho _{3}=\frac{F^{2}-\rho_{b0}}{a^{3}}\quad
,\quad \rho _{b}=\frac{\rho_{b0}}{a^{3}}\quad , \label{split}
\end{equation}
and
\begin{equation}
E= \frac{k}{3}\sqrt{\rho_{b0}}\quad ,\quad F=\sqrt{\rho_{b0}}%
B\quad . \label{barA}
\end{equation}
The individual equations of state of the different components are:
\begin{equation}
p_{1}=-\rho _{1}\quad ,\quad p_{2}=-\frac{1}{2}\rho _{2}\quad
,\quad p_{3}=0\quad ,\quad p_{b}=0\quad . \label{decompp}
\end{equation}
Consequently, our system is equivalent to a mixture of two dark
energy type fluids, one dark matter fluid and the baryon
component. While the BV model is a unified description of dark
matter/energy, the option of a decomposition may be useful in
comparing the scenario with some observational data. The X-ray
measurement of galactic clusters \cite{bartlett} is an example
which requires a separation of a matter component of the cosmic
medium. For a Chaplygin gas, different decompositions have been
proposed in the literature \cite {bertolamibis,winfried}.
Moreover, the decomposition (\ref{decomp}) with (\ref{split}) -
(\ref{decompp}) reveals, that the background dynamics of our model
differs from that of the $\Lambda$CDM model (including a baryon
component) by the existence of the second dark energy component
with energy density $\rho _{2}$.

The relevant cosmological quantities are the present density parameters  $%
\Omega _{v0}$ and $%
\Omega _{b0}$ which represent the fractions of the viscous fluid
and of the pressureless matter, respectively, with respect to the
total density (remember that we restrict ourselves to the flat
case), the Hubble parameter today $H_{0}$, the present value of
the deceleration parameter $q_{0}$, and the age of the universe
$t_{0}$. The apparently new parameter of the present model, the
viscosity coefficient, can be expressed in terms of the other
quantities. In fact, the relation (\ref{viscous-dens}) can be
rewritten in a more convenient way. Using  Friedmann's equation
today, we have
\begin{eqnarray}
\label{def1} H_0^2 = \frac{8\pi G}{3}(\rho_{b0} + \rho_{v0}) \quad
&\Rightarrow& \quad \Omega_{v0}
+ \Omega_{b0} = 1 \quad ,\\
\label{def2} \Omega_{b0} = \frac{8\pi G}{3H_0^2}\rho_{b0} \quad
&,& \quad \Omega_{v0} = \frac{8\pi G}{3H_0^2}\rho_{v0} \quad .
\end{eqnarray}
Combining (\ref{viscous-dens}) with (\ref{def1}) and (\ref{def2})
we obtain the relation
\begin{equation}
B+\frac{k}{3}=\frac{1}{\sqrt{\Omega _{b0}}}\quad .
\end{equation}
Now, if we use equation (\ref{em5}) with $p = p_{v} =-3\xi _{0}\frac{\dot{a}}{%
a}$ and the definition of the deceleration parameter $q_{0}=-\frac{\ddot{a}a%
}{\dot{a}^{2}}$, we find
\begin{equation}
-2q_{0}+1=\frac{24\pi G\xi _{0}}{H_{0}}\quad .
\end{equation}
From the definition of $k$ in equation (\ref{defk}) we have
\begin{equation}
k=\frac{24\pi G\xi _{0}}{H_{0}}\frac{1}{\sqrt{\Omega _{b0}}}\quad
.
\end{equation}
Using equation (\ref{em5}), evaluated today, it follows that
\begin{equation}
k=\frac{1-2q_{0}}{\sqrt{\Omega _{b0}}}\quad ,
\end{equation}
which leads to
\begin{equation}
B=\frac{2}{3}\frac{1+q_{0}}{\sqrt{\Omega _{b0}}}\quad .
\label{Bq0}
\end{equation}
Hence, the viscous fluid energy density becomes,
\begin{equation}
\rho _{v}=\rho _{b}\biggr\{f^{2}(a)-1\biggl\}\quad ,\quad f(a)=\frac{1}{3%
\sqrt{\Omega
_{b0}}}\biggr[2(1+q_{0})+(1-2q_{0})a^{3/2}\biggl]\quad .
\label{def3}
\end{equation}
All these relations will be useful for defining the observables
that will be constrained later on by the SNe Ia and matter power
spectrum data.

\section{A dynamical system analysis}
\label{dynamical system}

The relations of the previous section that will be used to compare
the theoretical predictions of our model with observational data,
are valid under the restriction $\nu = 0$. It is important to have
an idea of how serious this restriction is. If the case $\nu = 0$
is very particular, our results will not be conclusive. Therefore
it is desirable to have at least a qualitative analysis of the
general case (\ref{em1})-(\ref{em4}) which, because of its
complexity, cannot be solved analytically. Such an analysis could
reveal whether or not the case $\nu = 0$ is, in a sense, typical.
To this purpose we shall use dynamical system techniques for a
system of non-linear differential equations \cite{dynsyst}.

The system of equation (\ref{em1})-(\ref{em4}) can be recast in a
more convenient form by using geometric unities $8 \pi G = 1$ and
fixing the scale so that $\xi_0 = 1$. Defining $x = \frac{\dot
a}{a}$ and $y = \rho_v$, the following system of equations is
obtained:
\begin{eqnarray}
\dot x &=& -\frac{3}{2}x \left(x - y^\nu \right) = P(x, y) \label{dotx}\\
\dot y &=& - 3xy \left(1 - 3 xy^{\nu - 1}\right)= Q(x, y)\ .
\label{doty}
\end{eqnarray}
Since the density $\rho_{b} = 3x^2 - y$ of the baryonic matter
must be non-negative, the solutions that have physical meaning are
those which satisfy $y \leq 3 x^2$.

\begin{figure}[!t]
\begin{minipage}[t]{0.46\linewidth}
\includegraphics[width=\linewidth]{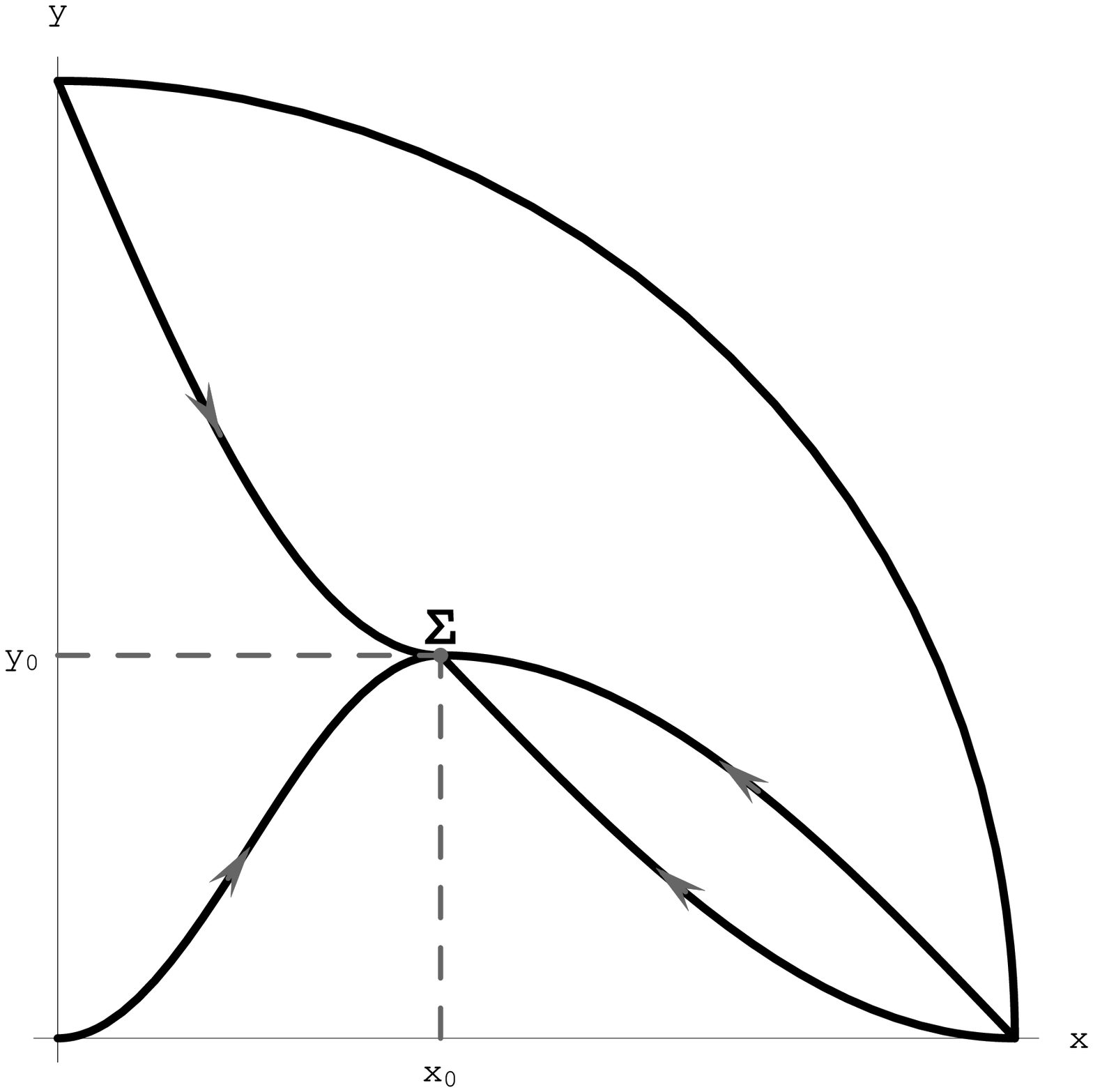}
\end{minipage} \hfill
\begin{minipage}[t]{0.46\linewidth}
\includegraphics[width=\linewidth]{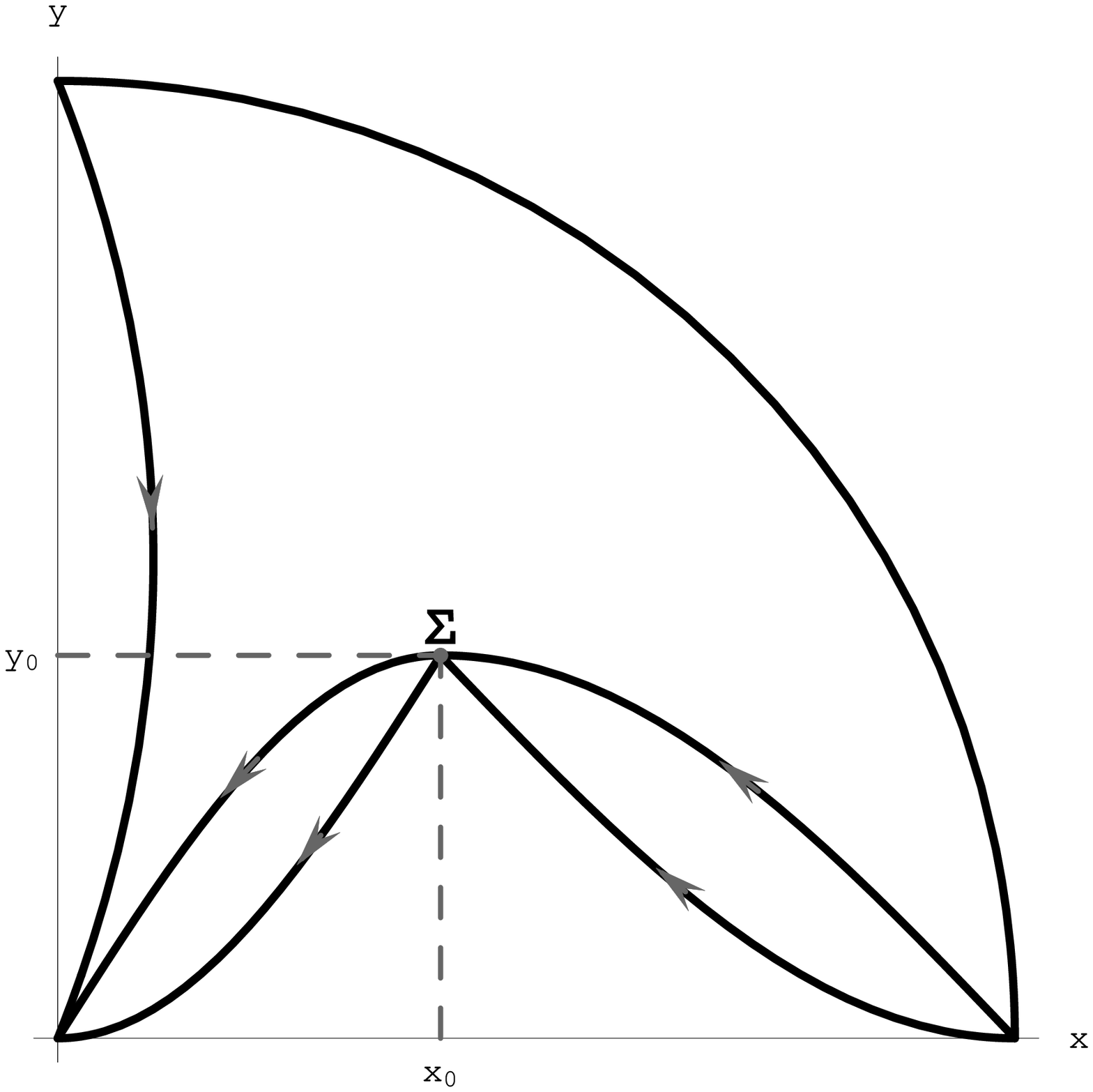}
\end{minipage} \hfill
\caption{{\protect\footnotesize Phase diagrams for $\protect\nu <
1/2$ (upper pane) and for $\protect 1/2 < \nu < 3/2$ (lower
panel). For $\protect \nu >  3/2$ the directions of the arrows in
the lower panel have to be reversed.}} \label{fig1}
\end{figure}

In a first step we have to identify the critical points
$(x_0,y_0)$ for which $P(x_0,y_0) = Q(x_0,y_0) = 0$. These points
are either attractors or repellers or saddle points. The nature of
a critical point is determined by the eigenvalues of the matrix
\begin{center}
$%
\bordermatrix{& & \cr&P_x(x_0,y_0)&P_y(x_0,y_0)\cr
&Q_x(x_0,y_0)&Q_y(x_0,y_0)\cr} $ ,
\end{center}
where $P_{x^i} = \frac{\partial P}{\partial x^i}$, $x^i = x, y$
and $Q_{x^i} = \frac{\partial Q}{\partial x^i}$. In general, there
are two
eigenvalues $%
\lambda_{1,2}$ for the characteristic equation of this matrix. If
both eigenvalues are positive, than the critical point is
repeller; if both are negative, the critical point is an
attractor; if one is positive and the other one is negative, than
the critical point is a saddle. With the help of a suitable
transformation \cite{dynsyst}, the critical points at infinity can
be studied. A projection onto the Poincar\'e sphere \cite{dynsyst}
will then allow us  to represent the entire phase space in a
finite region in the $x-y$ plane.

It turns out that the entire $Oy$ axis is a stationary solution
for the system, representing a static space-time.
A singular point, denoted by $%
\Sigma $, is $\left(x_0, y_0 \right) =
\left(3^{\frac{\nu}{1-2\nu}}, 3^{\frac{1}{1 - 2 \nu}} \right)$
where $\nu \neq \frac{1}{2}$. The characteristic
equation for this critical point is given by $\lambda^2 + \alpha_0 \left(%
\frac{3}{2} - \nu \right) \lambda + \alpha_0^{2} \left(
\frac{1}{2}- \nu \right) = 0$ where $\alpha_0 =
3^{\frac{1-\nu}{1-2\nu}}$. The eigenvalues are $\lambda_{1} =
\left(\nu - \frac{1}{2}\right)\alpha_0$ and $\lambda_{2} =
-\alpha_0$. The phase diagram depends on the value of the
parameter $\nu$. If $\nu < 1/2$, the critical point $\Sigma$ is an
attractor: all solutions converge to it. This point represents the
de Sitter phase. Hence, all solutions approach a de Sitter phase
asymptotically.

If $1/2 < \nu < 3/2$, the point $\Sigma$ becomes a saddle point
and all solutions approach a Minkowski space-time in the origin.
If $\nu > 3/2$, $\Sigma$ is still a saddle point, but the
directions of the curves change: all solutions leave the Minkowski
space-time. These different behaviors are shown in figure
(\ref{fig1}). For the point $\Sigma$ we have $3 x_{0}^{2} =
y_{0}^{2}$, which corresponds to a vanishing baryon density.
(Recall that the range $y > 3 x^{2}$ is unphysical.) The
deceleration parameter $q$ can be written as $q = - 1 -
\frac{\dot{H}}{H^{2}}$, equivalent to $q = - 1 -
\frac{\dot{x}}{x^{2}}$. Combination with Eq.~(\ref{dotx}) shows
that there is accelerated expansion $q < 0$ for $x < 3 y^{\nu}$.
The special case $\nu = 0$ has $\left(x_0, y_0 \right) = \left(1,
3 \right)$ and accelerated expansion for $x < 3$. This solution is
of the type of the curves that approach the point $\Sigma$ from
right below in the left panel of Fig.~\ref{fig1}. Notice that
there is a projection effect due to the circumstance that infinity
is mapped onto one single point on the $x$ axis. The curve that
points from the origin towards $\Sigma$ in the left part of
Fig.~\ref{fig1} corresponds to a phantom scenario with $\rho_{b} =
0$ which separates the physical and unphysical regions. We are not
interested here in this type of models.

\section{Constraining the model using SNe Ia data}
\label{SN}

The type Ia supernovae allow us to obtain informations about the
universe at high redshifts. Today, observers have detected about
300  of these objects with redshifts up to almost $2$. In what
follows, we will use the more restricted sample of 182 SNe Ia of
the Gold06 data set which consists of objects for which the
observational data are of very high quality \cite{riess2006}. The
relevant quantity for our analysis is the moduli distance $\mu$,
which is obtained from the luminosity distance $D_{L}$ by the
relation
\begin{equation}
\mu =5\log\biggr(\frac{D_{L}}{Mpc}\biggl)+25\quad .
\end{equation}

\begin{table}[!t]
\begin{center}
\begin{tabular}{|c|c|c|c|c|}
\hline\hline
&  &  &  &  \\[-7pt]
& Bulk & Generalized & Traditional & $\Lambda$CDM \\
& Viscosity & Chaplygin gas & Chaplygin gas &  \\[2pt] \hline
&  &  &  &  \\[-7pt]
$\chi^{2}$ & $162.71$ & $157.52$ & $157.88$ & $160.07$ \\[2pt] \hline
&  &  &  &  \\[-7pt]
$B$ & $1.81$ & $-$ & $-$ & $-$ \\[2pt] \hline
&  &  &  &  \\[-7pt]
$\alpha$ & $-$ & $1.90$ & $1$ & $0$ \\[2pt] \hline
&  &  &  &  \\[-7pt]
$\bar{E}$ & $-$ & $0.897$ & $0.825$ & $1$ \\[2pt] \hline
&  &  &  &  \\[-7pt]
$\Omega_{b0}$ & $0.0550$ & $0.0523$ & $0.0530$ & $0.054$ \\[2pt] \hline
&  &  &  &  \\[-7pt]
$\Omega_{dm0}$ & $-$ & $-$ & $-$ & $0.284$ \\[2pt] \hline
&  &  &  &  \\[-7pt]
$H_{0}$ & $62.38$ & $63.97$ & $63.56$ & $62.8695$ \\[2pt] \hline
&  &  &  &  \\[-7pt]
$t_{0}$ & $15.92$ & $13.83$ & $14.09$ & $14.83$ \\[2pt] \hline
&  &  &  &  \\[-7pt]
$q_{0}$ & $-0.364$ & $-0.775$ & $-0.672$ & $-0.493$ \\[2pt] \hline
\end{tabular}
\end{center}
\caption{The best-fitting parameters, i.e., when
$\protect\chi^{2}$ is minimum, for the BV model, the GCG model,
the
traditional Chaplygin gas model ($\alpha = 1$) and the $\Lambda $%
CDM model ($\alpha = 0$) for a flat Universe ($\Omega _{k0}=0$%
). $H_{0}$ is given in $%
km/M\!pc.s$, $\bar{E} \equiv \frac{E}{E + F}$   in units of
$c^{2}$ ($c$ - speed of light) and $t_{0}$ in $Gy$. }
\label{tableBestFitSNeIa}
\end{table}

The computation of the luminosity distance follows the standard
procedure \cite {SN,colistete}. First, we recall its definition
\cite{weinberglivro,coles},
\begin{equation}
D_{L}=\frac{r}{a}=(1+z)r\quad ,
\end{equation}
where $r$ is the comoving coordinate of the source and $z$ is the
redshift, $z=-1+\frac{1}{a}$. The coordinate $r$ can be obtained
by considering the propagation of light  $ds^{2}=0$. This implies,
\begin{equation}
ds^{2}=c^{2}dt^{2}-a^{2}dr^{2}=0\quad \Rightarrow \quad r=-c\int_{1}^{a}%
\frac{da}{a\dot{a}}\quad .
\end{equation}
From Friedmann's equation one obtains,
\begin{equation}
\dot{a}=\sqrt{\frac{\Omega _{m0}}{a}}H_{0}\,f(a)\quad .
\end{equation}
In terms of the redshift $z$, the luminosity distance may be
expressed as
\begin{equation}
D_{L}=3(1+z)\frac{c}{H_{0}}\int_{0}^{z}\frac{dz^{\prime }}{%
2(1+q_{0})(1+z^{\prime })^{3/2}+(1-2q_{0})}\quad ,
\end{equation}
or, using $B$ and $\Omega _{m0}$,
\begin{equation}
D_{L}=(1+z)\frac{c}{H_{0}}\int_{0}^{z}\frac{d\,z^{\prime
}}{1+\sqrt{\Omega _{m0}}B\ ((z^{\prime }+1)^{3/2}-1)}\quad .
\end{equation}

\begin{table}[!t]
\begin{center}
\begin{tabular}{|c|c|c|c|c|}
\hline\hline
&  &  &  &  \\[-7pt]
& Bulk & Generalized & Traditional & $\Lambda$CDM \\
& Viscosity & Chaplygin gas & Chaplygin gas &  \\[2pt] \hline
&  &  &  &  \\[-7pt]
$B$ & $1.81_{-0.34}^{+0.38}$ & $-$ & $-$ & $-$ \\[2pt] \hline
&  &  &  &  \\[-7pt]
$\alpha$ & $-$ & $1.25_{-1.78}^{+4.24}$ & $1$ & $0$ \\[2pt] \hline
&  &  &  &  \\[-7pt]
$\bar{E}$ & $-$ & $0.943_{-0.265}^{+0.056}$ &
$0.825_{-0.074}^{+0.057}$ & $1$
\\[2pt] \hline
&  &  &  &  \\[-7pt]
$\Omega_{b0}$ & $0.055_{-0.011}^{+0.011}$ & $0.053_{-0.010}^{+0.011}$ & $%
0.053_{-0.010}^{+0.011}$ & $0.054_{-0.010}^{+0.011}$ \\[2pt] \hline
&  &  &  &  \\[-7pt]
$\Omega_{dm0}$ & $-$ & $-$ & $-$ & $0.284_{-0.072}^{+0.082}$ \\[2pt] \hline
&  &  &  &  \\[-7pt]
$H_{0}$ & $62.34_{-1.72}^{+1.75}$ & $63.74_{-2.07}^{+1.93}$ & $%
63.45_{-1.79}^{+1.76}$ & $62.80_{-1.75}^{+1.76}$ \\[2pt] \hline
&  &  &  &  \\[-7pt]
$t_{0}$ & $15.85_{-0.91}^{+1.08}$ & $13.72_{-0.62}^{+1.21}$ & $%
14.04_{-0.05}^{+0.57}$ & $14.80_{-0.62}^{+0.73}$ \\[2pt] \hline
&  &  &  &  \\[-7pt]
$q_{0}$ & $-0.357_{-0.123}^{+0.130}$ & $-0.816_{-0.133}^{+0.362}$ & $%
-0.671_{-0.087}^{+0.109}$ & $-0.502_{-0.103}^{+0.136}$ \\[2pt] \hline
&  &  &  &  \\[-7pt]
$p(q_{0}<0)$ & $6.22\,\sigma$ & $100\,\%$ & $100\,\%$ & $100\,\% $ \\%
[2pt] \hline\hline
\end{tabular}
\end{center}
\caption{The estimated parameters for the BV model, the GCG model,
the traditional Chaplygin gas model and the $\Lambda $CDM model
for a flat Universe ($\Omega _{k0}=0$). We use the Bayesian
analysis to obtain the peak of the one-dimensional marginal
probability and the $2\,\protect\sigma $
confidence region for each parameter. $H_{0}$ is given in  km/s/M\!pc, $\bar{E%
}\equiv \frac{E}{E + F}$ in units of $c^{2}$ ($c$ - speed of
light) and $t_{0}$ in Gy. } \label{tableParEstSNeIa}
\end{table}

In order to compare the theoretical results with the observational
data, the first step is to compute the quality of the fitting
through the least square fitting quantity $\chi ^{2}$. We adopt
the HST (Hubble Space Telescope) prior \cite{freedman} for
$H_{0}$, as well as the cosmic nucleosynthesis prior for the
baryonic density parameter $\Omega _{b0}h^{2}$ (where $h$ is
$H_{0}$ divided by $100$ km/s/M\!pc), which leads to
\begin{equation}
\chi ^{2}=\sum_{i}\frac{\left( \mu _{0,i}^{o}-\mu _{0,i}^{t}\right) ^{2}}{%
\sigma _{\mu
_{0},i}^{2}}+\frac{(h-0.72)^{2}}{0.08^{2}}+\frac{(\Omega
_{b0}h^{2}-0.0214)^{2}}{0.0020^{2}}\quad .  \label{Chi2HSTOmegab}
\end{equation}
The quantities $\mu _{0,i}^{o}$ are the distance moduli,
observationally measured for each supernova of the $182$ Gold06
SNe Ia dataset \cite {riess2006} and the $\mu _{0,i}^{t}$ are the
corresponding theoretical values. The $\sigma _{\mu _{0},i}^{2}$
represent the measurement errors and include the dispersion in the
distance moduli, resulting from  the galaxy redshift dispersion
which is due to the peculiar velocities of the objects (cf.
\cite{riess,riess2006}).

The smaller the $\chi ^{2}$, the better the agreement between the
theoretical model and the observational data. Table
\ref{tableBestFitSNeIa} shows that the $\chi ^{2}$ value for BV
model is competitive with the GCG model (cf. Eq.~(\ref{rhoGCG})),
the traditional Chaplygin gas model (cf. Eq.~(\ref{rhoGCG})
with $\alpha = 1$) and the $%
\Lambda $CDM model. Compared with the $%
\Lambda $CDM and the (generalized) Chaplygin gas models, the age
of the Universe is greater for the bulk viscous model and its
deceleration parameter is less negative, i.e., the Universe is
less accelerated.

Using Bayesian statistics, a probability distribution can be
constructed from the $\chi^2$ parameter \cite{colistete}. In the
present case, there are three free parameters: the Hubble
parameter today $H_0$, the pressureless baryonic density parameter
$\Omega_{b0}$ and the auxiliary quantity $B$ which is connected to
the present value of the deceleration parameter (cf. (\ref{Bq0})).
Marginalization over one or two of these parameters will lead to
corresponding two- or one-dimensional representations. The details
of the Bayesian statistics and computational analysis, relying on
BETOCS (\textbf{B}ay\textbf{E}sian \textbf{T}ools for \textbf{O}%
bservational \textbf{C}osmology using \textbf{S}Ne Ia), are given
in ref. \cite{colistete}.

The main feature of the parameter estimation for the viscous fluid
model is the Gaussian-like distribution around the preferred
value.

\begin{figure}[!t]
\begin{minipage}[t]{0.46\linewidth}
\includegraphics[width=\linewidth]{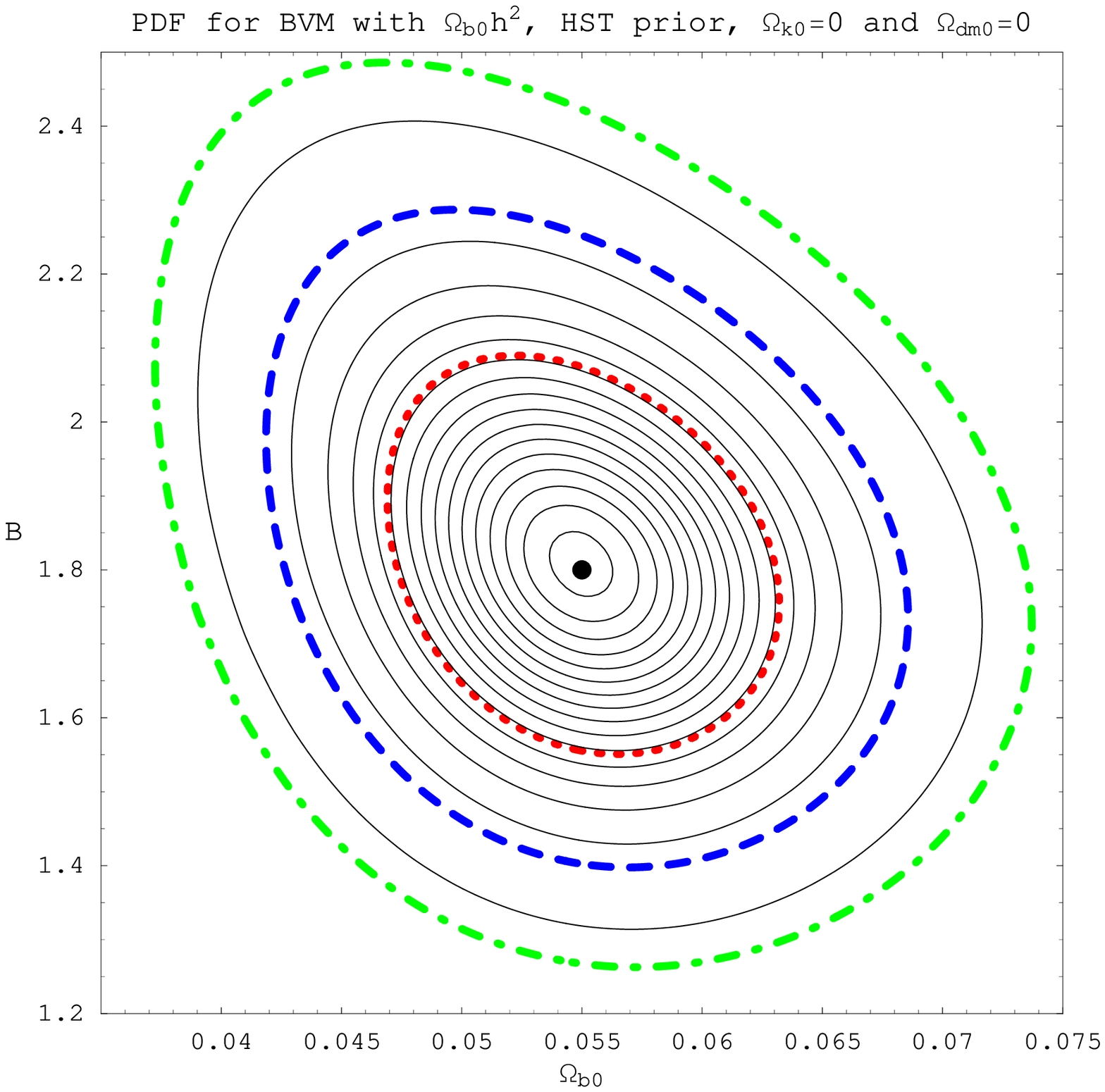}
\end{minipage} \hfill
\begin{minipage}[t]{0.46\linewidth}
\includegraphics[width=\linewidth]{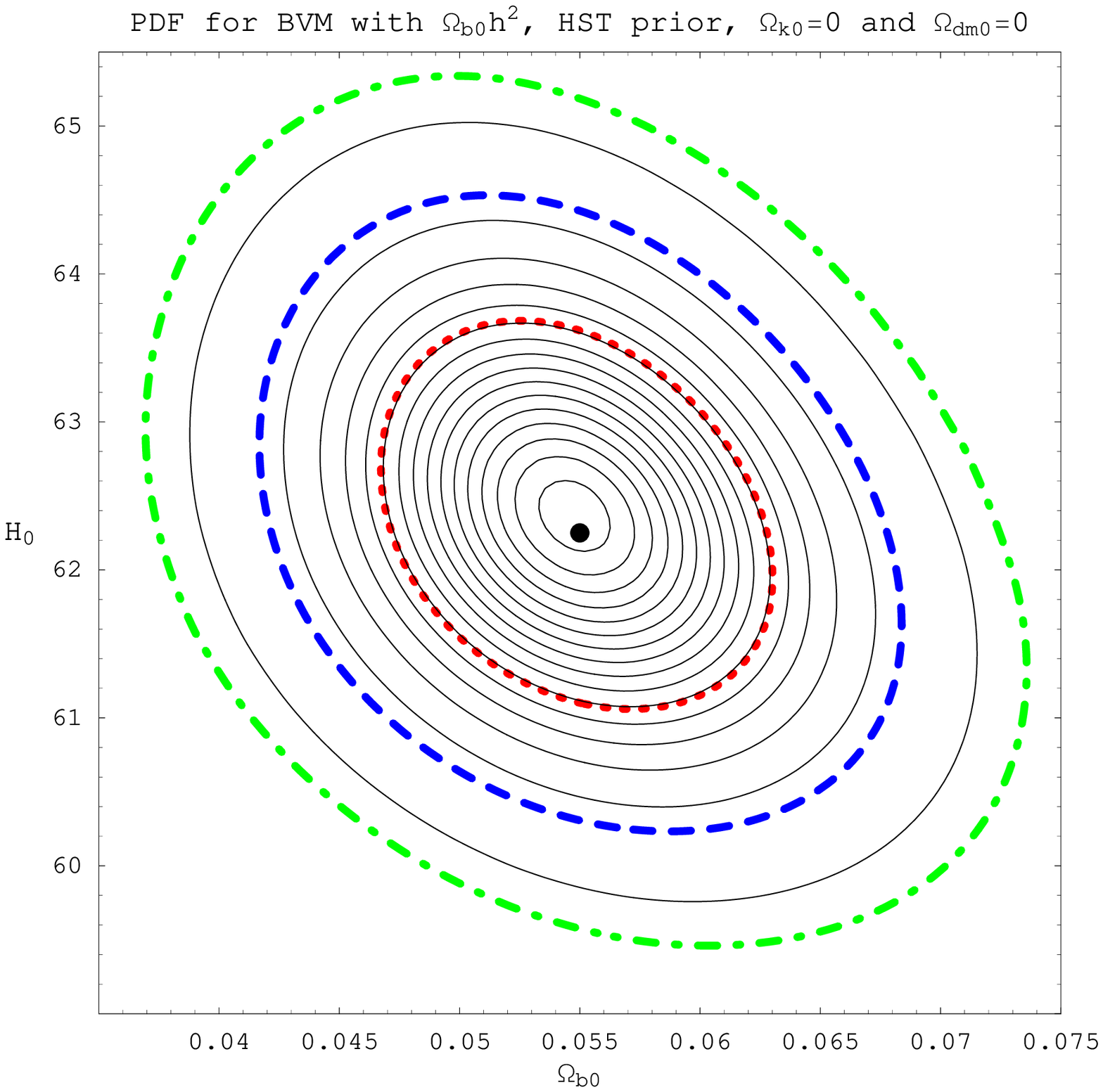}
\end{minipage} \hfill
\caption{{\protect\footnotesize The two-dimensional plots of the
Probability Density Function (PDF) as function for the parameter
$B$, the baryon density and the Hubble constant. The joint PDF
peak is shown by the large dot, the confidence
regions of $1\,\protect\sigma $ ($68,27\%$) by the red dotted line, the $2\,%
\protect\sigma $ ($95,45\%$) in blue dashed line and the
$3\,\protect\sigma $ ($99,73\%$) in green dashed-dotted line.}}
\label{figs2DSNeIa}
\end{figure}

\begin{figure}[!t]
\begin{minipage}[t]{0.46\linewidth}
\includegraphics[width=\linewidth]{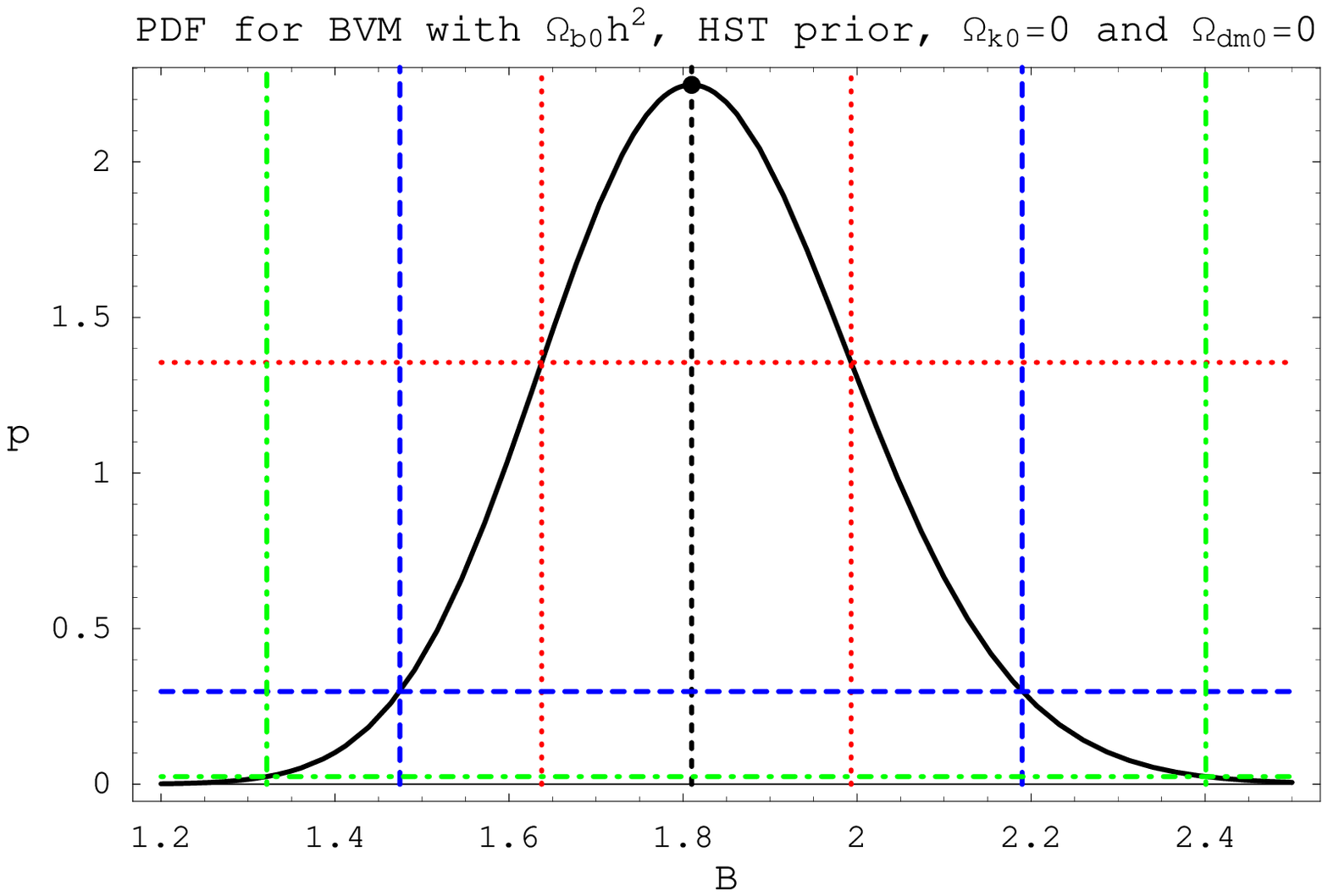}
\end{minipage} \hfill
\begin{minipage}[t]{0.46\linewidth}
\includegraphics[width=\linewidth]{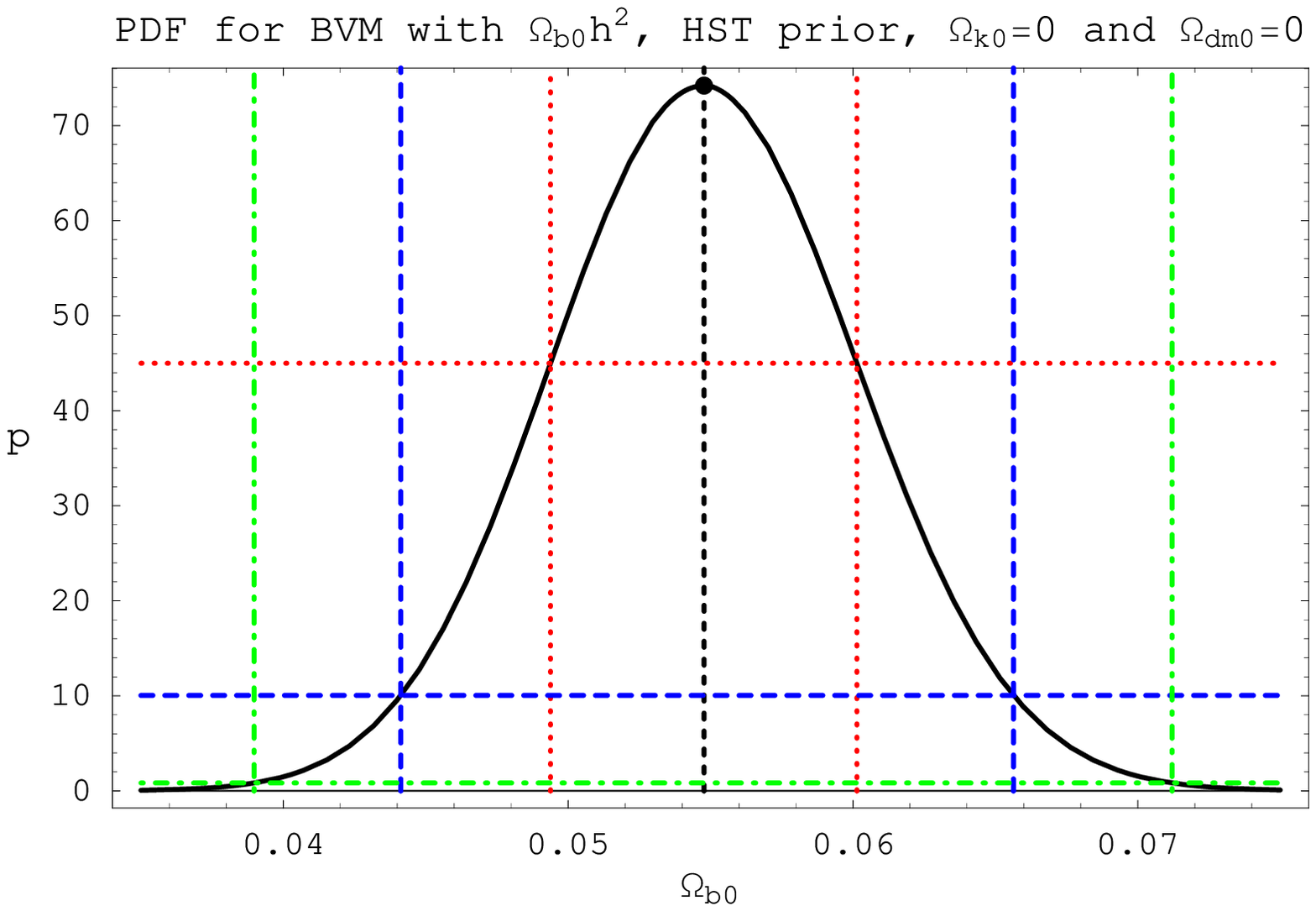}
\end{minipage} \hfill
\begin{minipage}[t]{0.46\linewidth}
\includegraphics[width=\linewidth]{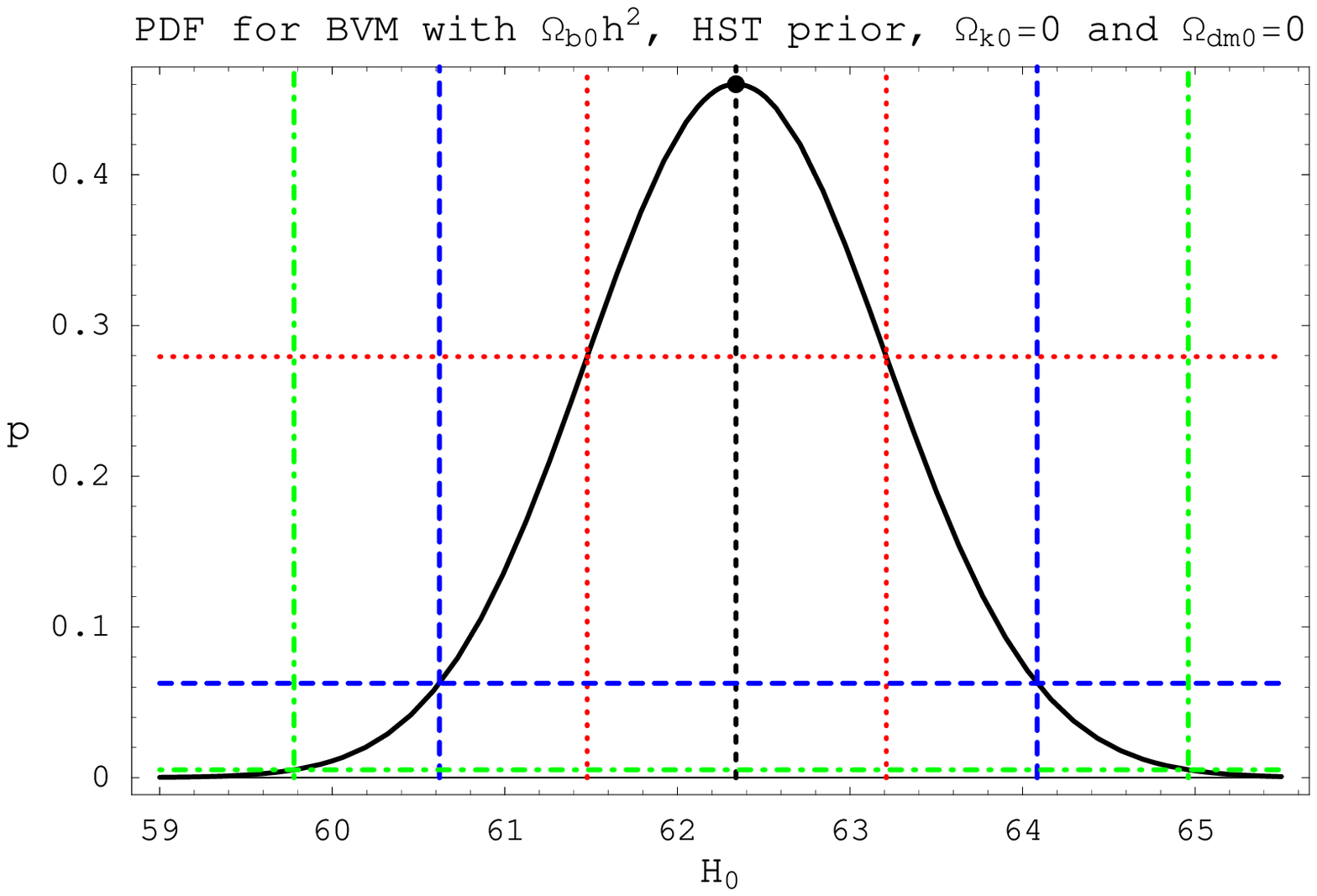}
\end{minipage} \hfill
\begin{minipage}[t]{0.46\linewidth}
\includegraphics[width=\linewidth]{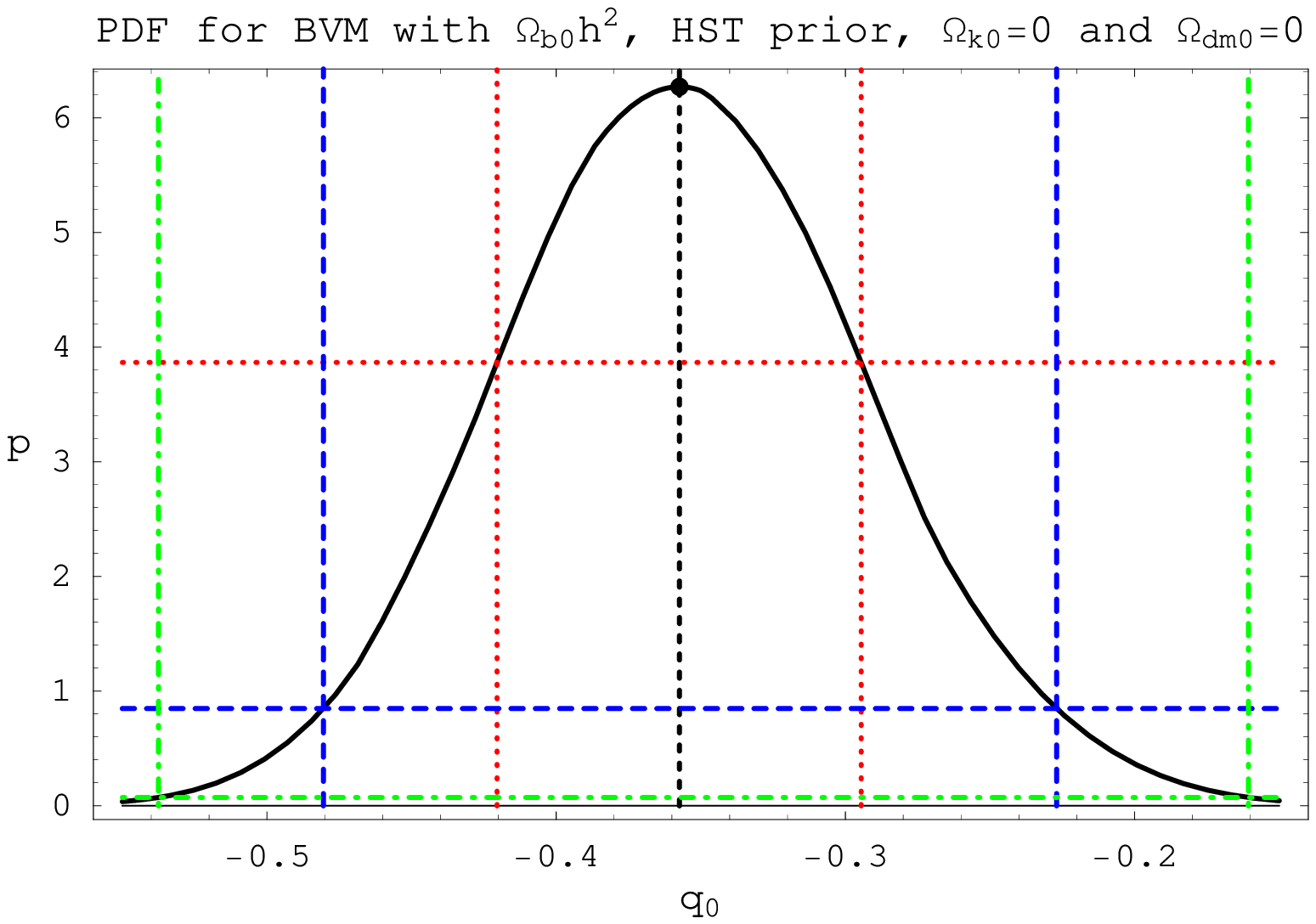}
\end{minipage} \hfill
\caption{{\protect\footnotesize The one-dimensional plots of the
PDF as function for the parameter $B$, the baryonic density, the
Hubble constant and the deceleration parameter. The joint PDF peak
is shown by the large dot, the confidence regions of
$1\,\protect\sigma $ ($68,27\%$) by the red dotted line, the
$2\,\protect\sigma $ ($95,45\%$) in blue dashed line and the
$3\,\protect\sigma $ ($99,73\%$) in green dashed-dotted line.}}
\label{figs1DSNeIa}
\end{figure}

This can easily be seen both in the table \ref {tableParEstSNeIa}
(almost symmetric error bars) and in the two and one-dimensional
diagrams of figures \ref{figs2DSNeIa} and \ref{figs1DSNeIa}. In
figure \ref{figs2DSNeIa} we show the two-dimensional probability
distribution for the parameters $B$, $\Omega _{b0}$ and $H_{0}$ at
one, two and three sigma, corresponding, respectively, to $68\%$,
$95\%$ and $99\%$ confidence levels. The corresponding
one-dimensional probability distributions are shown in figure
\ref{figs1DSNeIa}. The main features of the model are: the age of
the universe is considerably larger than for the $\Lambda$CDM and
GCG models, around $t_{0}\sim 16\,$ Gy, with a small dispersion;
the baryon density is around $0.05$, a little larger than
predicted by nucleosynthesis, but there is a marginal agreement;
the Hubble constant is around $62\,$ km/s/Mpc, i.e. much smaller
than the one predicted by the CMB, which is around $72\,$
km/s/Mpc, but in agreement with the $\Lambda$CDM and GCG models
when only the data from SNe Ia are used; the deceleration
parameter is bigger (smaller absolute value) than in those models.
For the parameter estimations in the $\Lambda$CDM and in the GCG
models, see ref. \cite{colistete} and references therein.

The Gaussian nature of the distribution is responsible for the
fact that the best fitting set of values is very close to the
marginalized parameter estimations, with a small dispersion.
Generally, this is neither true for the $\Lambda$CDM model nor for
the GCG model. Again: the most remarkable differences between the
viscous model and the $\Lambda$CDM model are the larger age of the
universe and the smaller absolute value of the deceleration
parameter.

\section{The matter power spectrum}
\label{matter power spectrum}

\subsection{The perturbed equations}

In order to write the first order perturbed equations, we
re-express the field equations as

\begin{eqnarray}
R_{\mu\nu} &=& 8\pi G \biggr\{T^v_{\mu\nu} - \frac{1}{2}g_{\mu\nu}T^v%
\biggl\} + 8\pi G \biggr\{T^b_{\mu\nu} - \frac{1}{2}g_{\mu\nu}T^b\biggl\} %
\quad, \\
{T_v^{\mu\nu}}_{;\mu} = 0 \quad &,& \quad T_v^{\mu\nu} = (\rho_v +
p_v)u^\mu
u^\nu - p_v g^{\mu\nu} \quad , \\
p_v &=&  - \xi_{0}{u^\mu}_{;\mu} \quad , \\
{T_b^{\mu\nu}}_{;\mu} = 0 \quad &,& \quad T_b^{\mu\nu} =
\rho_{b}u^\mu u^\nu \quad .
\end{eqnarray}
We introduce the quantities,
\begin{equation}
\tilde g_{\mu\nu} = g_{\mu\nu} + h_{\mu\nu} \quad , \quad
\tilde\rho = \rho + \delta\rho \quad , \quad \tilde u^\mu = u^\mu
+ \delta u^\mu \quad ,
\end{equation}
where $g_{\mu\nu}$, $\rho$ and $u^\mu$ are the known background
solutions for the
metric, the energy density and the four velocity, respectively, and $h_{\mu\nu}$%
, $\delta\rho$ and $\delta u^\mu$ are the corresponding
perturbations. We will perform the calculations using the
synchronous coordinate condition,
\begin{equation}
h_{\mu0} = 0 \quad .
\end{equation}
Since we will be interested mainly in perturbations that are
inside the horizon, the choice of the gauge is not really
essential.

The perturbed Ricci tensor takes the form,
\begin{equation}
\delta R_{\mu\nu} = {\chi^\rho_{\mu\nu}}_{;\rho} - {\chi^\rho_{\mu\rho}}%
_{\;\nu} \quad
\end{equation}
where
\begin{equation}
\chi^\rho_{\mu\nu} =
\frac{g^{\rho\sigma}}{2}\biggr[h_{\sigma\mu;\nu} +
h_{\sigma\nu;\mu} - h_{\mu\nu;\sigma}\biggl] \quad ,
\end{equation}
are the perturbations of the Christoffel symbols. The relevant
non-vanishing components are:
\begin{eqnarray}
\chi^i_{0j} &=& - \frac{1}{2}\biggr(\frac{h_{ij}}{a^2}\biggl)^{\displaystyle\cdot} \quad , \\
\chi^0_{ij} &=& -\frac{\dot h_{ij}}{2} \quad , \\
\chi^k_{ij} &=& - \frac{1}{2a^2}\biggr\{h_{ik,j} + h_{ij,k} - h_{ij,k}%
\biggl\} \quad .
\end{eqnarray}
For the  components of the perturbed  energy-momentum tensor we
have
\begin{eqnarray}
\delta T^{00} &=& \delta\rho \quad , \\
\delta T^{0i} &=& (\rho + p)\delta u^i \quad , \\
\delta T^{ij} &=& h^{ij}p - g^{ij}\delta p \quad .
\end{eqnarray}
\\
Using the definitions,
\begin{equation}
h = \frac{h_{kk}}{a^2} \quad , \quad \Theta = \delta^i_{,i} \quad,
\quad
\delta_b = \frac{\delta\rho_{b}}{\rho_{b}} \quad , \quad \Delta p_v = \frac{%
\delta p_v}{\xi_0} \quad ,
\end{equation}
and performing a plane wave decomposition of all perturbed
functions according to
\begin{equation}
\delta f(\vec x, t) = \delta f(t)e^{i\vec k\cdot\vec x} \quad ,
\end{equation}
where $\vec k$ is the wave vector,  we find, after a long but
straightforward calculation, the following set of first order
perturbed equations:
\begin{eqnarray}
\ddot h + 2\frac{\dot a}{a}\dot h = 8\pi G(\delta\rho_v + 3\delta
p_v) +
8\pi G\rho_{b}\delta_b \quad , \\
\delta\dot\rho_v + 3\frac{\dot a}{a}(\delta\rho_v + \delta p_v) +
(\rho_v +
p_v)\biggr(\Theta - \frac{\dot{h}}{2}\biggl) = 0 \quad , \\
(\dot\rho_v + \dot p_v)\Theta + (\rho_v + p_v)\dot\Theta + 5\frac{\dot a}{a}%
(\rho_v + p_v)\Theta -\frac{k^2}{a^2}\delta p_v = 0 \quad , \\
\dot\delta_b = \frac{\dot h}{2} \quad .
\end{eqnarray}
Replacing $\dot{h}$ by using the last relation, we obtain
\begin{eqnarray}  \label{pea-4}
\ddot\delta_b + 2 \frac{\dot a}{a}\dot\delta_b - 4\pi
G\rho_{b}\delta_b - 4\pi
G\delta\rho_v - 12\pi G\xi_0\Delta p_v &=& 0\quad , \\
\delta\dot\rho_v + 3\frac{\dot a}{a}(\delta\rho_v + \xi_0\Delta
p_v) -
(\rho_v + p_v)\Delta p_v &=& 0 \quad , \\
(\dot\rho_v + \dot p_v)\Theta + (\rho_v + p_v)\dot\Theta + 5\frac{\dot a}{a}%
(\rho_v + p_v)\Theta - \frac{k^2}{\xi_0}{a^2}\Delta p_v &=& 0 \quad , \\
\Theta - \dot\delta_b + \Delta p_v &=& 0 \quad .\label{last}
\end{eqnarray}
With the help of the constraint (\ref{last}), this system can be
further reduced to a set of three coupled differential equations:
\begin{eqnarray}
\ddot\delta_b + 2 \frac{\dot a}{a}\dot\delta_b - 4\pi
G\rho_{b}\delta_b - 4\pi
G\delta\rho_v - 12\pi G\xi_0\Delta p_v &=& 0\quad ; \label{3system1}\\
\delta\dot\rho_v + 3\frac{\dot a}{a}(\delta\rho_v + \xi_0\Delta
p_v) -
(\rho_v + p_v)\Delta p_v &=& 0 \quad ;\label{3system2} \\
(\dot\rho_v + \dot p_v)\dot\delta_b + (\rho_v + p_v)\ddot\delta_b + 5\frac{%
\dot a}{a}(\rho_v + p_v)\dot\delta_b &=&  \nonumber \\
(\dot\rho_v + \dot p_v)\Delta p_v + (\rho_v + p_v)\Delta\dot p_v + 5\frac{%
\dot a}{a}(\rho_v + p_v)\Delta p_v + \frac{k^2}{\xi_0}{a^2}\Delta
p_v & & \quad . \label{3system3}
\end{eqnarray}
\\
Now it is convenient to introduce the background equations and the
quantities in Eq.~(\ref{def3}) from which we find
\begin{eqnarray}
\frac{4\pi G\rho_{b}}{H_0^2} &=&
\frac{3}{2}\frac{\Omega_{b0}}{a^3} \quad ,
\label{background1} \\
\frac{4\pi G\rho_v}{H_0^2} &=& \frac{3}{2}\frac{\Omega_{b0}}{a^3}[f(a)^2 - 1%
] \quad, \label{background2}\\
\frac{12\pi G\xi_0}{H_0} &=& \frac{1}{2}(1 - 2q_0) \quad .
\label{background3}
\end{eqnarray}
The following relations will be useful later on:
\begin{eqnarray}
\frac{p_v}{\rho_v} &\equiv& g(a) = - \frac{1}{3}\frac{f(a)}{f(a)^2 - 1}\frac{%
1 - 2q_0}{\sqrt{\Omega_{b0}}}a^{3/2} \quad , \\
\dot\rho_v &=& - 3\frac{\dot a}{a}(\rho_v + p_v) \quad , \\
\dot p_v &=& \frac{1 - 2q_0}{2}(\rho_v + \rho_{b} + p_v) \quad .
\end{eqnarray}
\\
In a next step we divide the set of equations (\ref{3system1}) -
(\ref{3system3}) by $H_0^2$ and redefine $%
H_0^{-1}\Delta p_v \rightarrow \Delta p_v$. Again, after a fairly
long calculation, the perturbed equations reduce to the following
system of coupled linear differential equations:
\begin{eqnarray}
\ddot\delta_b + 2\frac{\dot a}{a}\dot\delta_b - \frac{3}{2}\frac{\Omega_{b0}%
}{a^3}\delta - \frac{3}{2}\frac{\Omega_{b0}}{a^3}(f^2 - 1)\delta_v - \frac{1%
}{2}(1 - 2q_0)\Delta p_v &=& 0\quad , \\
\dot\delta_v - 3\frac{\dot a}{a}g\delta_v - (1 + 2g)\Delta p_v &=&
0\quad ,
\\
(1 + g)\ddot\delta_b + \biggr\{2\frac{\dot a}{a}(1 + g) + \frac{1 - 2q_0}{2}%
\biggr[\frac{f^2}{f^2 - 1} + g\biggl]\biggl\}\dot\delta_b &=&  \nonumber \\
(1 + g)\Delta\dot p_v + \biggr\{2\frac{\dot a}{a}(1 + g) + \frac{1 - 2q_0}{2}%
\biggr[\frac{f^2}{f^2 - 1} + g\biggl] & &  \nonumber \\
+ \biggr(k\, l_H\biggl)^2\frac{(1 - 2 q_0)a}{9\Omega_{b0}(f^2 - 1)}%
\biggl\}\Delta p_v & & \quad ,
\end{eqnarray}
where $l_H = c/H_0$.

Now, we re-express the perturbed equations in terms of the
variable $a$ according to (\ref{prime}), which implies
\begin{equation}
\dot\delta = \delta^{\prime}\dot a \quad , \quad \ddot\delta =
\dot a^2\delta^{\prime\prime}+ \ddot a\delta^{\prime}\quad .
\end{equation}
We also need the following relations, which result from the
background equations:
\begin{equation}
\dot a^2 = \Omega_{b0}\frac{f^2}{a} \quad , \quad \frac{\ddot
a}{\dot a^2} = - \frac{1}{2a}\biggr\{3g\frac{f^2 - 1}{f^2} +
1\biggl\} \quad .
\end{equation}
With the redefinition $\Delta_v = \frac{\Delta p_v}{\dot a}$, the
final form for the set of perturbed equations is:
\begin{eqnarray}
\label{f-pe1} \delta_b'' + \biggr\{- \frac{3}{2}g\frac{f^2 -
1}{f^2 a} + \frac{3}{2a}\biggl\}
\delta_b' - \frac{3}{2}\frac{\delta_b}{f^2 a^2} &=& \nonumber\\
\frac{3}{2}\frac{f^2 - 1}{f^2 a^2}\delta_v - \frac{3}{2}\frac{g}{a}
\frac{f^2 - 1}{f^2}\Delta_v & &;\\
\delta'_v - 3\frac{g}{a}\delta_v - (1 + 2g)\Delta_v &=& 0 \quad ; \\
\label{f-pe3} (1 + g)\delta_b'' + \biggr\{-
\frac{3}{2}\frac{g}{a}(1 + 2g)
\frac{f^2 - 1}{f^2} + \frac{3}{2a}\biggl\}\delta_b' &=& \nonumber \\
(1 + g)\Delta_v' + \biggr\{- \frac{3}{2}\frac{g}{a}(1 +
2g)\frac{f^2 - 1}{f^2} + \frac{3}{2a} - \biggr(k\,
l_H\biggl)^2\frac{g}{3f^2\Omega_{b0}}\biggl\}\Delta_v & &.
\end{eqnarray}

\subsection{Numerical integration}

Evidently, equations (\ref{f-pe1})-(\ref{f-pe3}) are too
complicated to admit analytical solutions. Hence, we proceed
integrating these equations numerically. In order to do so, an
important problem is to fix the initial conditions. A scale
invariant primordial spectrum (as predicted by the inflationary
scenario) is assumed. But, since we are interested in the power
spectrum today, we have to follow the evolution of this spectrum
until the present phase.

The corresponding procedure for the $\Lambda$CDM model has been
carried out in terms of the  BBKS transfer function \cite
{bardeen,sugiyama,martin}. This function is characterized by
\begin{eqnarray}
\tilde{q}=\tilde{q}(k)=%
\frac{k}{(h \Gamma)\,\mathrm{Mpc}^{-1}}\,,\quad
\Gamma\,=\,\Omega_{M}^0\,h\,
e^{-\,\Omega_B^0\,-\,\left(\Omega_B^0/\Omega_{M}^0\right) } \, ,
\label{tq}
\end{eqnarray}
where $\Omega_{M}^0$ and $\Omega_{B}^0$ are the density parameters
of dark matter and baryonic matter of the $\Lambda$CDM model. The
baryonic transfer function is approximated by by the numerical fit
\cite{bardeen}:
\begin{eqnarray}  \label{jtf}
T(k)=\frac{\,\mbox{ln}\, (1+2.34 \tilde{q})}{2.34 \,\tilde{q}}\,
\left[1+3.89 \tilde{q} + \left(16.1 \tilde{q}\right)^2 +
\left(5.46 \tilde{q}\right)^3
+ \left(6.71 \tilde{q}\right)^4\right]%
^{-1/4}\,.
\end{eqnarray}
The quantity of interest is the power spectrum for the baryonic
matter today. This spectrum is given by
\begin{equation}  \label{Powers}
P(k) \,=\, |\delta_b(k)|^2 \,=\, A\,\,k\,\,T^2(k)\,\frac{g^2(\Omega_T^0)}{%
g^2(\Omega^0_{M})}\,,
\end{equation}
where $\Omega_T^0$ is the total density parameter of the
$\Lambda$CDM model. The function $g$ in (\ref{Powers}) is the
growth function
\begin{eqnarray}  \label{CPT}
g(\Omega)=\frac{5\Omega}{2}\, \left[\Omega^{4/7} - \Omega_{\Lambda} + \Big(1+%
\frac{\Omega}{2}\Big)
\Big(1+\frac{\Omega_\La}{70}\Big)\right]^{-1}\,
\end{eqnarray}
in which $\Omega_{\Lambda}$ represents the fraction of the total
energy, contributed by the cosmological constant. The
normalization coefficient $A$ in (\ref{Powers}) can be fixed by
using the COBE measurements of the CMB anisotropy spectrum. This
coefficient is connected to the quadrupole momentum
\cite{sugiyama,martin} $Q_{rms}$ of this spectrum by the relation
\begin{eqnarray}
A \,=\, (2l_H)^4\, \frac{6\pi^2}{5}\,\frac{Q^2_{rms}}{T_0^2} \,,
\label{AA}
\end{eqnarray}
where $T_0 = 2.725\pm0.001$ is the present CMB temperature. The
quadrupole anisotropy is taken as
\begin{equation}
Q_{rms} = 18 \mu\,K \,.
\end{equation}
This value is obtained from the COBE normalization and is
consistent with the more recent results of the WMAP measurements
which use the prior of a scale invariant spectrum
\cite{martin,hinshaw}. Taking all these estimates into account,
one may fix
\begin{eqnarray}
A = 6.8\times 10^{5}\,h^{-4} \,\mathrm{Mpc}^4\,,  \label{A}
\end{eqnarray}
which corresponds to the initial vacuum state for the density
perturbations used in the BBKS transfer function.

We recall that all the relations (\ref{tq})-(\ref{A}) are valid
for the $\Lambda$CDM model. Strictly speaking, we would have to
redo all the calculations to obtain the analogue of the expression
(\ref{Powers}) for our bulk viscous model. However, since we are
interested in the shape of the spectrum, there exists a simpler
way which relies on the circumstance that the growth function $g$
in (\ref{CPT}) is a pure background quantity. Use of the
$\Lambda$CDM growth function and the $\Lambda$CDM  initial values
also for our model will result in an overall shift of the spectrum
$P$ without affecting the structure of the latter. The resulting
data will then be a fit to a ``wrong", i.e. unphysical
$\Lambda$CDM model, in which the composition of the dark sector
differs from the standard one with roughly $30\%$ of dark matter
and $70\%$ of a cosmological constant. However, this procedure
provides us with the correct structure of the spectrum for our
bulk viscous model. Keep in mind that the background dynamics of
our model differs from that of the $\Lambda$CDM model by the
existence of a second dark energy component (component 2) in
Eq.~(\ref{split}).

Well inside the matter dominated phase (say, $z\sim 500$), we use
the $\Lambda$CDM initial conditions, insert them in equations
(\ref{f-pe1})-(\ref{f-pe3}), and let this system evolve. Then we
compute the power spectrum $P(k)=\delta _{b}^{2}(k)$ at $z=0$,
equivalent to $a=1$. Finally, we compare the theoretical results
with the power spectrum obtained by the 2dFGRS
observational program. The spectrum depends essentially on two parameters: $%
\Omega_{b0}$ and $q_{0}$. From now on, relying on the results of
the type Ia SNe analysis and the primordial nucleosynthesis
constraint, we fix $\Omega _{b0}=0.04$ (according to the type Ia
SNe analysis of the previous section, this is marginally satisfied
at least at $2-\sigma $).

\begin{figure}[!t]
\begin{minipage}[t]{0.46\linewidth}
\includegraphics[width=\linewidth]{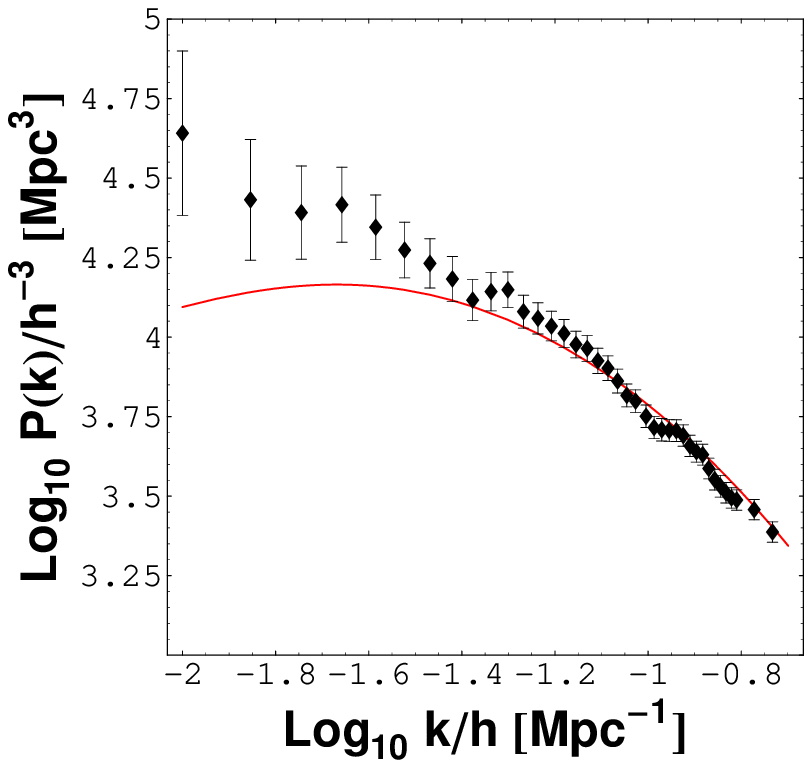}
\end{minipage} \hfill
\begin{minipage}[t]{0.46\linewidth}
\includegraphics[width=\linewidth]{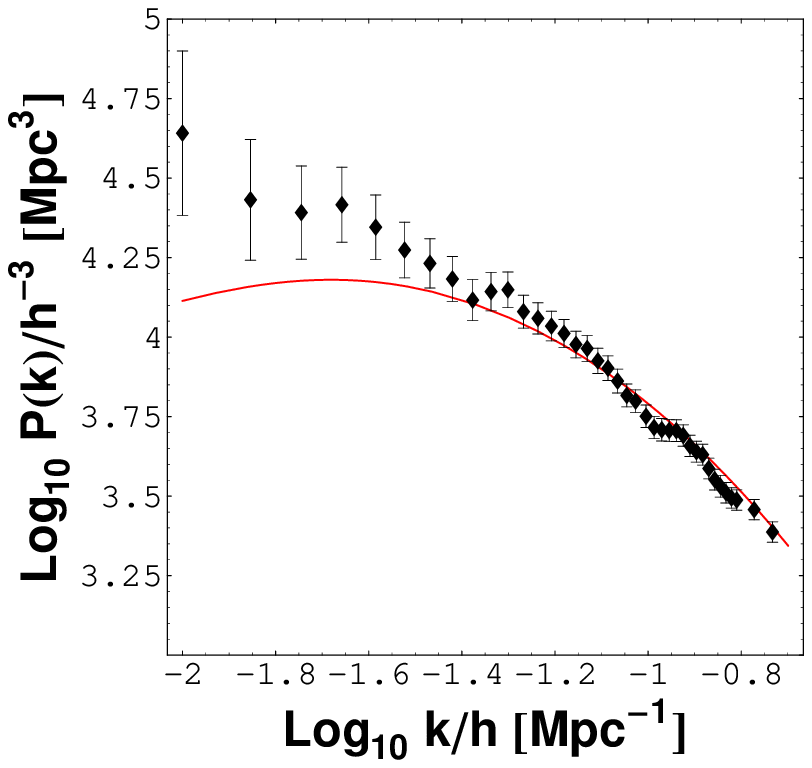}
\end{minipage} \hfill
\caption{{\protect\footnotesize Predicted power spectrum for the
dark viscous fluid compared with the 2dFGRS observational data for
$q_0 = -0.3$ (left) and $q_0 = -0.4$ (right). The initial
conditions using the BBKS transfer function correspond to ``wrong"
ratios (see text) of dark matter and dark energy: $\Omega_{dm0} =
0.49$ and $\Omega_{\Lambda0} = 0.51$ (left) and $\Omega_{dm0} =
0.52$ and $\Omega_{\Lambda0} = 0.48$ (right). }} \label{figpower1}
\end{figure}

\begin{figure}[!t]
\begin{minipage}[t]{0.46\linewidth}
\includegraphics[width=\linewidth]{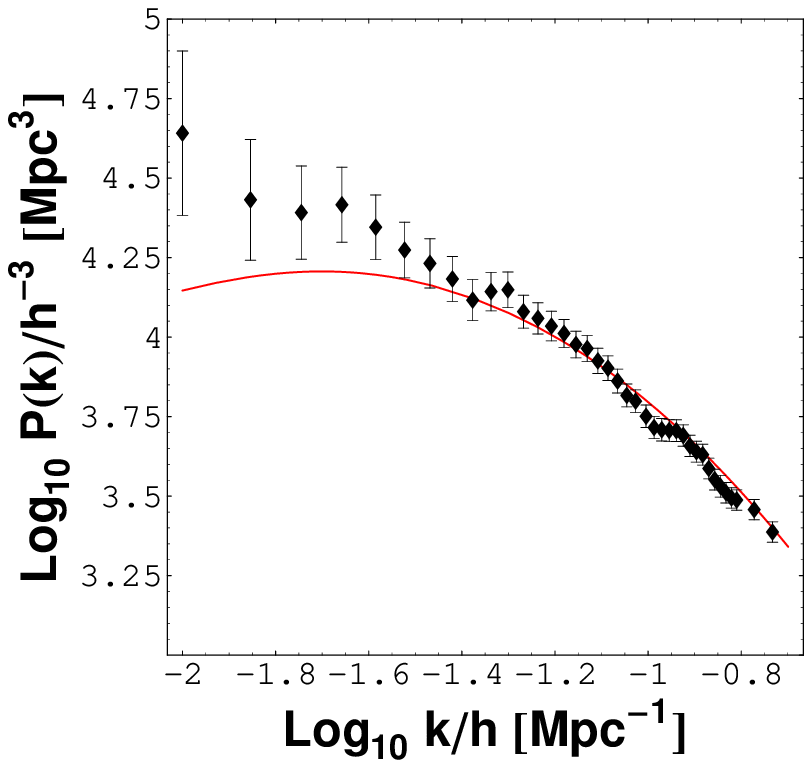}
\end{minipage} \hfill
\begin{minipage}[t]{0.46\linewidth}
\includegraphics[width=\linewidth]{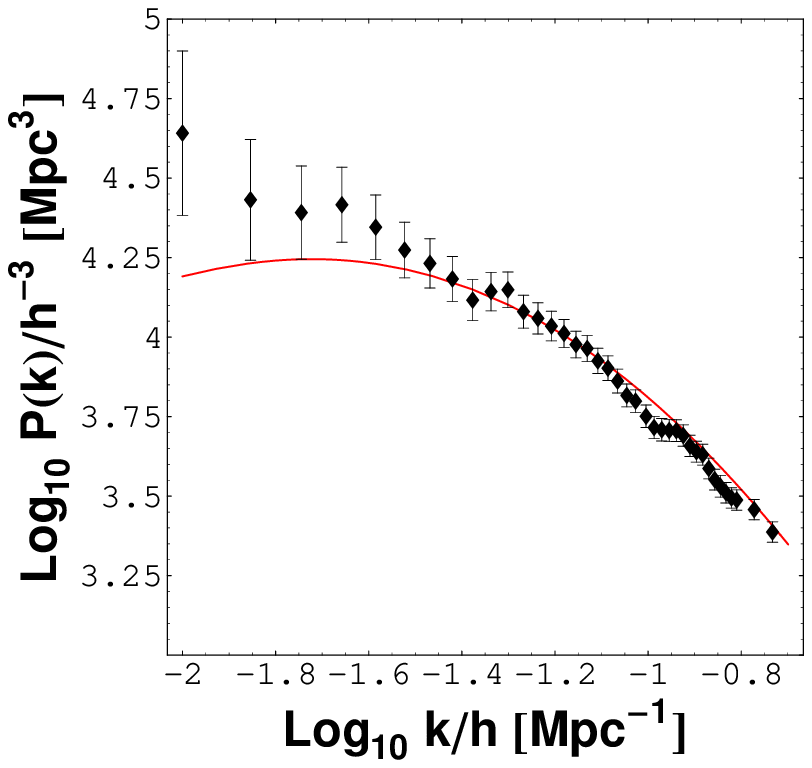}
\end{minipage} \hfill
\caption{{\protect\footnotesize Predicted power spectrum for the
dark viscous fluid compared with the 2dFGRS observational data for
$q_0 = -0.5$ (left) and $q_0 = -0.6$ (right). The initial
conditions using the BBKS transfer function correspond to ``wrong"
ratios of dark matter and dark energy: $\Omega_{dm0} = 0.44$ and
$\Omega_{\Lambda0} = 0.56$ (left) and $\Omega_{dm0} = 0.52$ and
$\Omega_{\Lambda0} = 0.48$ (right).}} \label{figpower2}
\end{figure}

In figures \ref{figpower1} and \ref{figpower2} we present
the best fitting for $%
q_{0}=-0.3,-0.4,-0.5,-0.6$. To have a good agreement, the ratio
between the quantity of dark matter/dark energy in the ``wrong"
$\Lambda$CDM reference model
has to be in the range from $1$ for $q_{0}=-0.3$, to about $2/3$ for $%
q_{0}=-0.6$. It is seen that the baryonic matter power spectrum of
the viscous fluid model is in good agreement with the
observational data for values of $q_{0}$ that are close to those
predicted by the type Ia SNe analysis, that is, $q_{0}\sim -0.4$.
We remark \textit{en passant} that a variation of the value of
$\Omega _{b0}$ within the interval predicted by the type Ia SNe at
$2-\sigma $ does not change substantially the results presented in
figures \ref{figpower1} and \ref{figpower2}. In any case, there is
no instability in the computed power spectrum. This is true not
only for the baryonic fluid power spectrum but also for dark
viscous fluid power spectrum. We note, however, that as $q_{0}$
approaches zero, there is a power depression at large scales
(small $k$) compared with the $\Lambda$CDM reference model.

\subsection{A physical interpretation}

In the homogeneous and isotropic background, the BV model is
equivalent to a GCG with $\alpha = - 1/2$. The qualitative
differences between both models at the perturbative level can be
traced back to a difference in the pressures that characterize the
cosmic medium in each of these cases. In the background both
pressures coincide. But while the perturbations for the GCG are
entirely adiabatic, this is not the case for our present model.

In the background we have $p = p_{v} = - 3 H \xi_{0}$ for the
viscous fluid and $p= - E \rho^{1/2}$ for the GCG with $\alpha = -
1/2$ (cf. (\ref{pGCG})). Use of Friedmann's equation $3H^{2} =
8\pi G \rho$ (for the spatially flat case) reveals that in both
cases we have an equation of state $p \propto - \rho^{1/2}$, where
the constants are related by $E = \xi_{0}\sqrt{24 \pi G}$. The
first order pressure perturbations for the GCG are
\begin{equation}
\delta p = \frac{\dot{p}}{\dot{\rho}}\delta \rho
\qquad\qquad\qquad (\mathrm{GCG\ model })
 \, ,
\end{equation}
where
\begin{equation}
\frac{\dot{p}}{\dot{\rho}} = - \alpha \frac{p}{\rho}
\qquad\qquad\qquad (\mathrm{GCG\ model })
 \,
\end{equation}
is the adiabatic sound speed square. For $\alpha > 0$ the sound
speed square is positive, giving rise to a (non-observed)
oscillatory behavior. For $\alpha < 0$ (our case) this square is
negative which results in instabilities which are not observed
either.

For the viscous fluid the pressure perturbations are qualitatively
different. In particular, the propagation of perturbations is no
longer given by the adiabatic sound speed. The perturbations are
non-adiabatic. The non-adiabaticity is characterized by
\begin{equation}
\delta p - \frac{\dot{p}}{\dot{\rho}}\delta \rho = -
\dot{p}\left(\frac{\delta \rho}{\dot{\rho}} - \frac{\delta
H}{\dot{H}}\right) \qquad\qquad\qquad (\mathrm{BV\ model })
 \, ,
 \label{nad}
\end{equation}
where $\delta H \equiv \delta \left(u^{\mu}_{;\mu}\right)/3$. In
general, the right hand side of (\ref{nad}) does not vanish (For a
similar situation see \cite{essay}). It is this non-adiabatic
character of the perturbations which makes the power spectrum well
behaved on small scales in contrast to the GCG case. A more
detailed analysis of this feature will be given elsewhere.

\section{Conclusions}
\label{conclusions}

We presented a two-component cosmological model in which one
component represents the dark sector, the other one pressureless
baryons. The dark sector is described by a single bulk viscous
fluid with a constant bulk viscosity coefficient $\xi = \xi_{0}=
$constant. A dynamical system analysis was used to show that the
assumption of a constant $\xi$ does not seem to be too restrictive
and that this model embodies quite general features and does not
represent a very particular configuration. In the homogeneous and
isotropic background the total energy density (i.e., including the
baryons) is equivalent to that of a GCG with $\alpha = -1/2$. But
while the perturbation dynamics of a GCG is entirely adiabatic,
the present model exhibits non-adiabatic features. As a
consequence, it can avoid the problems that usually plague such
type of unified dark matter/dark energy models, namely the
appearance of (non-observed) small scale oscillations or
instabilities. The predictions of our model were compared with SNe
Ia data and with the 2dFGRS results. For the comparison with SNe
Ia data we employed the so-called gold sample of 182 good quality
high redshift supernovae. Our Bayesian statistics analysis leads
to the following parameter evaluation at $2\sigma $ level:
$q_{0}=-0.357_{-0.123}^{+0.130}$, $\Omega
_{b0}=0.055_{-0.011}^{+0.011}$, $H_{0}=62.34_{-1.72}^{+1.75}\,$
km/s/M\!pc and $t_{0}=15.85_{-1.08}^{+0.91}\ $ Gy. The main
differences in comparison with a similar analysis for the
$\Lambda$CDM and the GCG models \cite{colistete} are: the age of
the universe is considerably larger; the deceleration parameter is
larger (smaller in absolute value); the baryon density is only
marginally compatible with the primordial nucleosynthesis result.
A remarkable feature is the (almost) perfect Gaussian distribution
for all parameters with a quite small dispersion.

To compare the results of our perturbation analysis with the
observed matter power spectrum we have used the BBKS transfer
function to fix the initial condition at a very high redshift
($z\sim 500$), deep in the matter dominated phase. Since the BBKS
transfer function has been conceived for a two-component
description of the dark sector, dark matter and a cosmological
constant, this must be seen with some caution. We have argued
that, in spite of this problem, the resulting power spectrum
preserves its shape and can be compared with observations. A good
fitting can be obtained for values of $q_{0}$ which are in
agreement with the type Ia SNe analysis. The best results are
found for $-0.3\gtrsim q_{0}\gtrsim -0.6$. Qualitatively, our
analysis confirms previous results for a simplified model of the
cosmic medium, in which the baryon component was neglected
\cite{rose}.

\acknowledgments{J.C.F. thanks the IMSP (B\'{e}nin) for the warm
hospitality during part of the elaboration of this work. He thanks
also CNPq (Brazil) and the CAPES/COFECUB French-Brazilian
scientific cooperation for partial financial support. W.Z.
acknowledges support by grants 308837/2005-3 (CNPq) and 093/2007
(CNPq and FAPES).}

\end{document}